\documentstyle[12pt]{article}

\begin{document}

\setcounter{page}{1}

\pagestyle{plain} \vspace{1cm}
\begin{center}
\Large{\bf Evolution of Perturbations in a Noncommutative Braneworld Inflation}\\
\small \vspace{1cm} {\bf Kourosh Nozari$^{a,b,*}$}\quad and\quad {\bf Siamak Akhshabi$^{a,\dagger}$}\\
\vspace{0.5cm} {\it $^{a}$Department of Physics,
Faculty of Basic Sciences,\\
University of Mazandaran,\\
P. O. Box 47416-95447, Babolsar, IRAN\\
\vspace{0.25cm} $^{b}$ Research Institute for Astronomy and
Astrophysics of Maragha,\\
P. O. Box 55134-441, Maragha, IRAN\\
\vspace{0.25cm}
$^{*}$knozari@umz.ac.ir\\
$^{\dagger}$s.akhshabi@umz.ac.ir}

\end{center}
\vspace{1.5cm}
\begin{abstract}
Following our previous work in noncommutative braneworld inflation (
Ref. [18]), in this letter we use smeared, coherent state picture of
noncommutativity to study evolution of perturbations in a
noncommutative braneworld scenario. We show that in this setup, the
early stage of the universe evolution has a phantom evolution with
imaginary effective sound speed. We show also that the amplitude of
perturbations in the commutative regime decays faster than the
noncommutative regime with the same parameter values, and as a
result we need smaller number of e-folds in the noncommutative
regime to
have a successful braneworld inflation.\\
{\bf PACS}: 02.40.Gh,\, 11.10.Nx,\, 04.50.-h, \,98.80.Cq\\
{\bf Key Words}: Inflation, Perturbations, Noncommutativity
\end{abstract}
\vspace{2cm}
\newpage
\section{Introduction}
Inspired by some aspects of string theory and loop quantum gravity,
\emph{fuzziness} of spacetime can be expressed using the following
relation for non-commutativity of coordinate operators[1,2]
\begin{equation}
[\hat{x}^i,\hat{x}^j]=i\theta^{ij}
\end{equation}
where $\theta^{ij}$ is a real, antisymmetric matrix, with the
dimension of length squared which determines the fundamental cell
discretization of spacetime manifold. As a consequence of above
relation, the notion of point in the spacetime manifold becomes
obscure as there is a fundamental uncertainty in measuring the
coordinates
\begin{equation}
\Delta x^{i}\Delta x^{j}\geq \frac{1}{2}|\theta^{ij}|.
\end{equation}
This finite resolution of the spacetime points especially affects
the cosmological dynamics in early stages of the universe evolution.
On the other hand, inflation has been identified as a great
opportunity to test theories of Planck scale physics including
noncommutative geometry. Essentially, effects of trans-Planckian
physics should be observable in the cosmic microwave background
radiation [3-9]. For this reason, various attempts to construct
noncommutative inflationary models have been done by adopting
various different approaches. These approaches include using
relation (1) for space-space [10] and space-time [11] coordinates
and constructing a noncommutative field theory on the spacetime
manifold by replacing ordinary product of fields by
Weyl-Wigner-Moyal $*-product$. Another way to incorporate effects of
high energy physics in inflationary models is using the generalized
uncertainty principle (GUP) which is a manifestation of the
existence of a fundamental length scale in the system [12].\\
Recently a new approach to noncommutative inflation has been
proposed by Rinaldi [13] using the coherent state picture of
noncommutativity introduced in [14]. This model is free from some of
the problems that plagued models based on $*-product$ such as
unexpected divergencies and UV/IR mixing (see [15] for a full
review). The key idea in this model is that noncommutativity
\emph{smears} the initial singularity and as a result there will be
an smooth transition between pre and post big bang eras via an
accelerated expansion. It has been shown that noncommutativity
eliminates point-like structures in the favor of smeared objects in
flat spacetime. As Nicolini {\it et al.}\, have shown [16] ( see
also [17] for some other extensions ), the effect of smearing is
mathematically implemented as a substitution rule: position
Dirac-delta function is replaced everywhere with a Gaussian
distribution of minimal width $\sqrt{\theta}$. In this framework,
they have chosen the mass density of a static, spherically
symmetric, smeared, particle-like gravitational source as follows
\begin{equation}
\rho_\theta(r)=\frac{M}{(2\pi\theta)^{\frac{3}{2}}}\exp(-\frac{r^2}{4\theta}).
\end{equation}
As they have indicated, the particle mass $M$, instead of being
perfectly localized at a point, is diffused throughout a region of
linear size $\sqrt{\theta}$. This is due to the intrinsic
uncertainty as has been shown in the coordinate commutators (1).

Recently we have constructed a noncommutative braneworld inflation
scenario [18] based on the idea that initial singularity is smeared
in a noncommutative background. Within the same streamline, the
purpose of this letter is to study time evolution of cosmological
perturbations in a braneworld inflation scenario in the context of
spacetime noncommutativity.

\section{Cosmological dynamics in the noncommutative RS II model}

The 5D field equations in the Randall-Sundrum (RS) II [19] setup are
\begin{equation}
^{(5)}\!G_{AB}=-\Lambda_5\,
^{(5)}\!g_{AB}+\delta(y)\,\frac{8\pi}{M_5^3}\left[ -\lambda
g_{AB}+T_{AB}\right],
\end{equation}
where $y$ is a Gaussian normal coordinate orthogonal to the brane (
the brane is localized at $y=0$), $\lambda$ is the brane tension,
and $T_{AB}$ is the energy-momentum tensor of particles and fields
confined to the brane . The effective field equations on the brane
are derived from the Gauss-Codazzi equations and junction conditions
(using $Z_2$-symmetry)[20,21]
\begin{equation}
G_{ab} = - \Lambda g_{ab} + \kappa^2 T_{ab} +
6\frac{\kappa^2}{\lambda} {\cal S}_{ab} - {\cal E}_{ab}\;,
\end{equation}
where ${\cal S}_{ab}\sim (T_{ab})^2$ is the high-energy correction
term, which is negligible for $\rho\ll\lambda$, while ${\cal
E}_{ab}$ is the projection of the bulk Weyl tensor on the brane. The
general form of the brane energy-momentum tensor for any matter
fields (scalar fields, perfect fluids, kinetic gases, dissipative
fluids, etc.), including a combination of different fields, can be
covariantly given in terms of a chosen 4-velocity $u^\mu$ as
\begin{equation}
T_{\mu\nu}=\rho u_\mu  u_\nu
+ph_{\mu\nu}+\pi_{\mu\nu}+q_{\mu}u_{\nu}+q_\nu  u_\mu \,.
 \label{3''}
\end{equation}
Here $\rho$ and $p$ are the energy density and isotropic pressure
and $q$ and $\pi$ are the momentum density and anisotropic stress
respectively. $h_{\mu\nu}$ defined as
\begin{equation}
h_{\mu\nu}=g_{\mu\nu}+u_\mu  u_\nu = \,  ^{(5)}\!g_{\mu\nu}-n_\mu
n_\nu +u_\mu  u_\nu
\end{equation}
projects into the comoving rest space at each event where $~n_{\nu}$
is the spacelike unit normal to the brane. The modified Friedmann
and Raychaudhuri equations in the background are [20]
\begin{equation}
H^2 = \frac{\kappa^2}{3} \rho\left(1+{\rho\over 2\lambda}\right)
+\frac{C}{a^{4}}+ \frac{1}{3} \Lambda - \frac{K}{a^2} \,,
\end{equation}
and
\begin{equation}
 \dot H= - {\kappa^2\over 2}(\rho+p)\left(1+ {\rho\over
\lambda}\right)-2{C\over a^4}+{K\over a^2}\,,
\end{equation}
respectively. By definition,
$C=\frac{\kappa^{2}}{3}\rho_{\varepsilon0}a^{4}_{0}$\, where
$\rho_{\varepsilon0}$ is the dark radiation energy density. For a
matter content consisted of a perfect fluid or a minimally coupled
scalar field the total effective energy density, pressure, momentum
density and anisotropic stress can be written as [21]
\begin{eqnarray}
\rho^{{\rm eff}} &=& \rho\left(1 +\frac{\rho}{2\lambda} +
\frac{\rho^{\varepsilon}}{\rho} \right)\;, \\  p^{\rm eff
} &=& p  + \frac{\rho}{2\lambda} (2p+\rho)+\frac{\rho^{\varepsilon}}{3}\;, \\
q^{\rm eff }_a &=& q_{a}^{\varepsilon}\;,
\\  \pi^{\rm eff}_{ab} &=&
\pi^{\varepsilon}_{ab}\,,
\end{eqnarray}
where superscript $\varepsilon$ denotes the contribution of the bulk
Weyl tensor which enters the modified friedmann equation as a
non-local dark radiation term. Using these definitions, the modified
Friedmann and Raychaudhuri equations can be rewritten as
\begin{eqnarray}
H^2 &=& \frac{\kappa^2}{3} \rho^{\rm eff} + \frac{1}{3} \Lambda +
\frac{K}{a^2} \,, \\ \dot H &=& -\frac{\kappa^2}{2}(\rho^{\rm eff}
+p^{\rm eff})+\frac{K}{a^2}\,.
\end{eqnarray}
The tracefree property of $\cal{E}^{\mu}_{\nu}$ in equation (5)
implies that the pressure obeys
$P^{\varepsilon}={1\over3}\rho^{\varepsilon}$. \\
The local conservation equations on the brane are [21]
\begin{eqnarray}
&&\dot{\rho}+\Theta(\rho+p)=0\,,\\ && D_a p+(\rho+p)A_a =0
\end{eqnarray}
where $\Theta$ is the volume expansion rate, which reduces to $3H$
in the FRW background ($H$ is the background Hubble rate), $A_a$ is
the 4-acceleration, and $D_a$ is the covariant derivative in the
rest space. The non-local conservation equations for the dark
radiation matter can be expressed as [21]
\begin{eqnarray}
&&\dot{\rho}^\varepsilon+{{4\over3}}\Theta{\rho^{\varepsilon}}+D^a{q^{\varepsilon}}=0
\\&& \dot{q}^{\varepsilon}_{a}+4H{q^{\varepsilon}_{a}}
+{{1\over3}}D_a{\rho^{\varepsilon}}+{{4\over3}}{\rho^{\varepsilon}}A_a
+D^b{\pi_{ab}^{\varepsilon}} = -{(\rho+p)\over\lambda} D_a \rho\,.
\end{eqnarray}
We now suppose that the initial singularity that leads to RS II
geometry afterwards, is smeared due to spacetime noncommutativity. A
newly proposed model for the similar scenario in the usual 4D
universe suggests that one could write the energy density as [13,18]
\begin{equation}
\rho(t)=\frac{1}{32\pi^{2}\theta^{2}}e^{-t^{2}/4\theta}\,.
\end{equation}
Note that we suppose that the universe enters the RS II geometry
immediately after the initial smeared singularity which is a
reasonable assumption (for instance, from a M-theory perspective of
the cyclic universe this assumption seems to be reliable, see Ref.
[22]). Using equation (20), and setting $\Lambda=0=K$, the Friedmann
equation (14) in noncommutative space could be rewritten as follows
\begin{equation}
H^{2}=\frac{\kappa^{2}}{3}\rho^{\rm eff}(t)
\end{equation}
where $\rho^{\rm eff}$ is given by equation (10). From equation (16)
one can find the effective noncommutative pressure using equation
(20) as
\begin{equation}
p=-\rho+\frac{t}{6\theta}e^{-t^{2}/8\theta}\,.
\end{equation}
So, the equation of state parameter will be
\begin{equation}
\omega=-1+\frac{16}{3}{\pi }^{2} \theta\,{t}e^{-t^{2}/8\theta}
\end{equation}
and the speed of sound is
\begin{equation}
c^{2}_{s}=\frac{\dot{p}}{\dot{\rho}}=\frac{-3t-64\theta^{2}
\pi^{2}e^{-t^{2}/8\theta}+32\theta\pi^{2}t^{2}e^{-t^{2}/8\theta}}{3t}\,.
\end{equation}
Using equations (10) and (11) we can find the \emph{effective}
equation of state and speed of sound. To do this end, we note that
there are constraints from nucleosynthesis on the value of
$\rho^{\varepsilon}$ so that
$\frac{\rho^{\varepsilon}}{\rho}\leq0.03$ at the time of
nucleosynthesis [23,24]. In this respect, we can neglect this
contribution to find

$$\omega^{\rm eff}={\frac {1}{192}}\,{{\rm e}^{-\frac{1}{8}\,{\frac
{{t}^{2}}{\theta}}}}
 \bigg[ -192\,{\pi }^{2}{\theta}^{2}\lambda+1024\,t{{\rm e}^{{
\frac {{-t}^{2}}{8\theta}}}}{\pi }^{4}{\theta}^{3}\lambda-3\,{{\rm
e}^{{\frac {{-t}^{2}}{8\theta}}}}+32\,t{{\rm e}^{{\frac {{-t}^{2}}
{4\theta}}}}{\pi }^{2}\theta \bigg]\times $$
$$\bigg[ \theta \left( {\frac {1}{ 64}}\, \left( 64\,{\pi
}^{2}{\theta}^{2}\lambda+{{\rm e}^{{ \frac {{-t}^{2}}{8\theta}}}}
\right) {\pi }^{-2}{\theta}^{-2}{\lambda}^{ -1} \right) \bigg]
^{2}\times$$
\begin{equation}
{\pi }^{-2}{\theta}^{-4}{\lambda}^{-1}
 \bigg[ {{\rm e}^{{\frac {{-t}^{2}}{8\theta}}}} \left( {\frac {1}{
64}}\, \left( 64\,{\pi }^{2}{\theta}^{2}\lambda+{{\rm e}^{{ \frac
{{-t}^{2}}{8\theta}}}} \right) {\pi }^{-2}{\theta}^{-2}{\lambda}^{
-1} \right)  \bigg]^{-1}
\end{equation}
which simplifies to the following equation for the high energy
regime ($\rho\gg\lambda$)
\begin{equation}
\omega^{\rm eff}\approx-1+\frac{32}{3}{\pi }^{2}
\theta\,{t}e^{-t^{2}/8\theta}\,.
\end{equation}
Similarly, the effective speed of sound in the high energy regime
will be
\begin{equation}
(c^{2}_{s})^{\rm eff}\approx \frac{16}{3}{\pi }^{2}
\theta\,{t}e^{-t^{2}/8\theta}+\frac{-3t-64\theta^{2}
\pi^{2}e^{-t^{2}/8\theta}+32\theta\pi^{2}t^{2}e^{-t^{2}/8\theta}}{3t}\,.
\end{equation}
Figure $1a$ shows the evolution of the equation of state parameter
and the effective equation of state parameter as given by equations
(23) and (25) respectively. As one can see from this figure, there
is a small variation in $\omega$ and $\omega^{eff}$ around the
smeared singularity. Figure $1b$ shows the evolution of the
effective speed of sound. It is obvious from this figure that in
$t>0$ and in the high energy noncommutative regime, $c_{s}$ is
imaginary. In this respect, the evolution of the universe in
the early, inflationary stage is a phantom evolution.\\

\begin{figure}[htp]
\begin{center}
\includegraphics{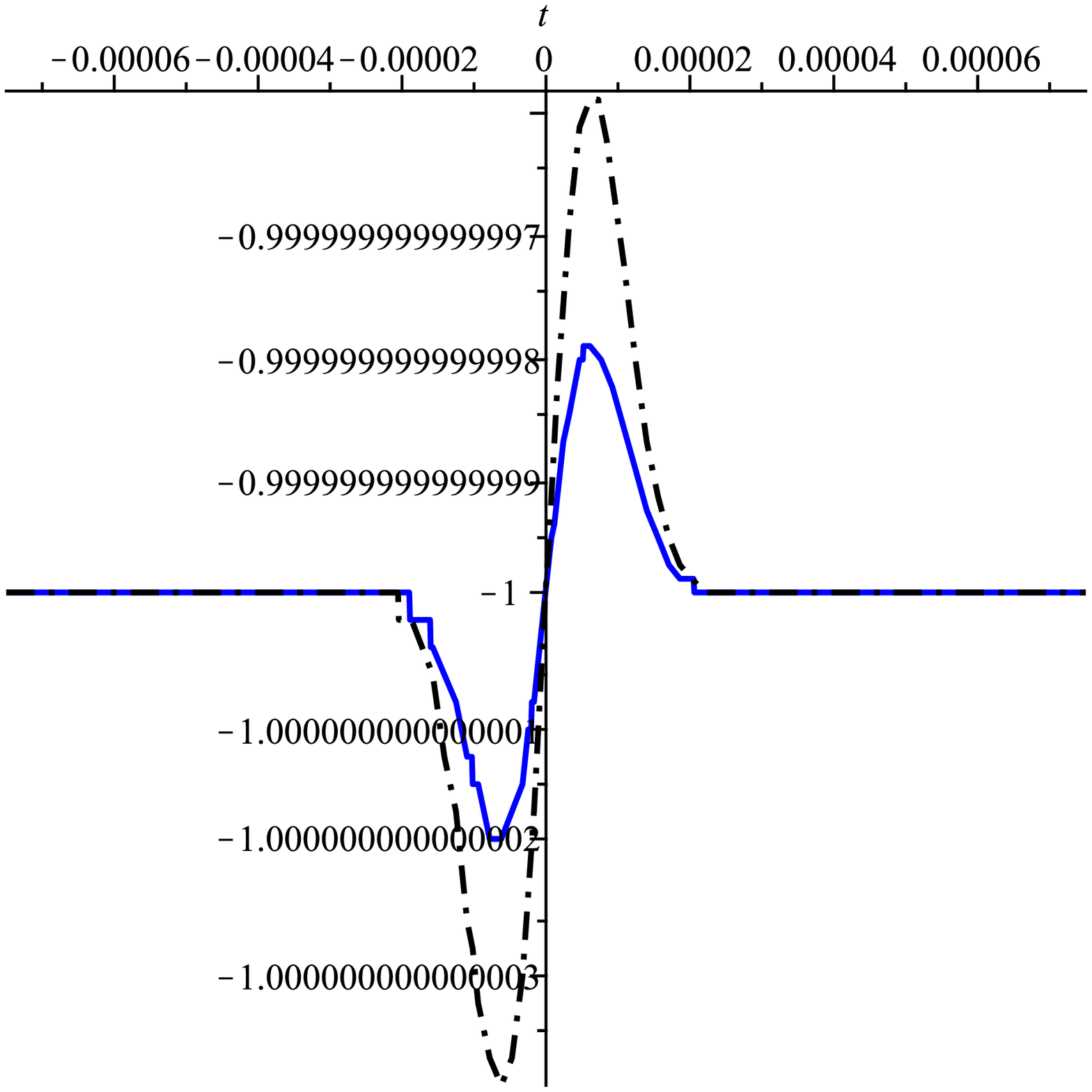} \vspace{5cm}\includegraphics{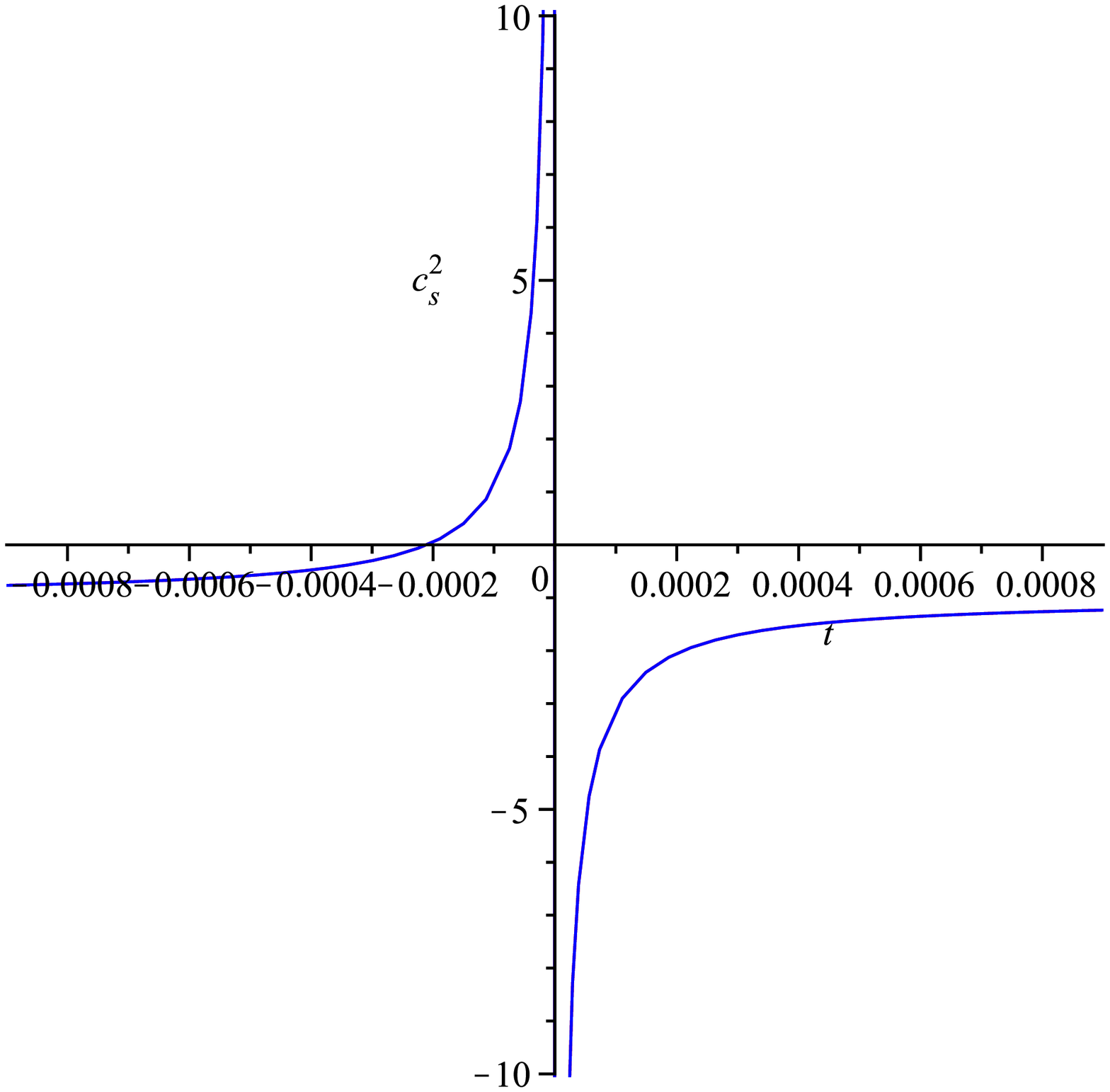}\vspace{2 cm}
\end{center}
\vspace{0cm} \caption{\small {a)  Evolution of the noncommutative
equation of state parameter (solid line) and the noncommutative
effective equation of state parameter ( dashed line) versus the
cosmic time. b) Evolution of the noncommutative effective speed of
sound versus the cosmic time. For $t>0$ and in the high energy
noncommutative regime, $c_{s}$ is imaginary ( a phantom
evolution).}}
\end{figure}\newpage

We use these results in the next section to determine time evolution
of the cosmological perturbations.

\section{Evolution of large scale scalar perturbations}
The evolution of cosmological perturbations in the Randall-Sundrum
braneworld  scenario has been studied extensively ( see for instance
[25] and references therein). To analyze the scalar perturbations in
our noncommutative setup, we define energy density and expansion
perturbations as [26] using the covariant 3+1 analysis developed in
[27]
\begin{equation}
\Delta={a^2\over\rho}D^2\rho\,,\quad\quad Z=a^2D^2\Theta\,.
\end{equation}
Similarly, the perturbations in nonlocal quantities associated with
the dark radiation matter are defined as
\begin{equation}
{U}={a^2\over\rho}D^2{\rho^\varepsilon}\,,\quad\quad
{Q}={a\over\rho} D^2 {q^{\varepsilon}}\,,\quad\quad {\Pi }={1\over
\rho}D^2{\pi^{\varepsilon}}\,.
\end{equation}
With these definitions, the equations governing on the evolution of
perturbations (equations (16)-(19)) will take the following forms
\begin{eqnarray}
\dot{\Delta} &=&3wH\Delta-(1+w)Z\,, \\  \dot{Z}
&=&-2HZ-\left({c_{\rm s}^2\over 1+w}\right) D^2\Delta-\kappa^2\rho {
U}-{{1\over2}}\kappa^2 \rho\left[1+ (4+3w){ {\rho\over\lambda}}-
\left({4c_{\rm
s}^2\over1+w}\right){\rho^\varepsilon\over\rho}\right] \Delta \,,\\
{ \dot{U}} &=& (3w-1)H{   U} + \left({4c_{\rm s}^2\over
1+w}\right)\left({{\rho^\varepsilon }\over\rho}\right) H\Delta
-\left({4{\rho^\varepsilon }\over3\rho}\right) Z-aD^2{ Q}\,,\\  {
\dot{Q}} &=&(3w-1)H{ Q}-{1\over3a}{ U}-{{2\over3}} a{
D^2\Pi}+{1\over3 a}\left[ \left({4c_{\rm s}^2\over
1+w}\right){{\rho^\varepsilon}\over\rho}-3(1+w) {
{\rho\over\lambda}}\right]\Delta\,.
\end{eqnarray}
where $\rho$\,, $\omega$ and $c_{s}$ are given by equations (20)\,,
(23) and (24) respectively. \\
In general, scalar perturbations on the brane cannot be predicted by
brane observers without additional information from the bulk because
there is no equation for $\dot{\Pi}$ in the above set of equations.
Nevertheless, it has been shown that on large scales one can neglect
the $D^{2}\Pi$ term in equation (33). So, on large scales, the
system of equations closes on the brane, and brane observers can
predict scalar perturbations from initial conditions intrinsic to
the brane without the need to solve the bulk perturbation
equations [26,27]. \\
To solve the above system of equations using the simplification
mentioned, we introduce two new variables; the first is a scalar
covariant curvature perturbation variable
\begin{equation}
C\equiv a^4D^2R = -4a^2HZ+2\kappa^2a^2\rho\left( 1+ {\rho \over
2\lambda} \right)\Delta+ 2\kappa^2a^2 \rho U\,,
\end{equation}
where $R$ is the Ricci curvature of the surfaces orthogonal to
$u^\mu$. The second variable is a covariant analog of the Bardeen
metric potential $\Phi_H$,
\begin{equation}
\Phi=\kappa^2a^2\rho \Delta\,.
\end{equation}
Along each fundamental world-line, covariant curvature perturbation,
$C$, is locally conserved
\begin{equation}
C=C_0\,,\quad \dot{C}_0=0\,.
\end{equation}
With these new variables, the system of equations reduces to
\begin{eqnarray}
\dot{\Phi}&=& -H\left[1+(1+w){\kappa^2\rho\over 2H^2}\left(1+
{\rho\over \lambda}\right)\right]\Phi  -
\left[(1+w){a^2\kappa^4\rho^2\over 2 H}\right]U +\left[(1+w)
{\kappa^2 \rho\over 4H}\right]C_0\\ \dot{U} &=&
-H\left[1-3w+{2\kappa^2{\rho^\varepsilon}\over 3H^2}\right]U -{2
{\rho^\varepsilon}\over 3 a^2 H\rho}\left[1+{\rho\over\lambda} - {6
c_{\rm s}^2H^2\over (1+w)\kappa^2\rho}\right]\Phi+
\left[{{\rho^\varepsilon}\over 3a^2H\rho}\right] C_0\,.
\end{eqnarray}
If there is no dark radiation in the background,
$\rho^\varepsilon=0$, then
\begin{equation}
U=U_0\exp\Big\{\int(3w-1)dN\Big\}\,,
\end{equation}
where $N$ is the number of e-folds. In this case, the above system
reduces to a single equation for $\Phi$ which is
\begin{eqnarray}
{d\Phi\over dN}+\left[1+{(1+w)\kappa^2\rho\over 2H^2}\left(1+
{\rho\over \lambda}\right)\right]\Phi =~{}\left[{(1+w)\kappa^2
\rho\over 4H^2}\right]C_o - \left[{3(1+w)a_o^2\rho^2\over \lambda
H^2}\right]e^{2N}U
\end{eqnarray}
where $U$  is given by (39). We use these results in the next
section to study noncommutative modifications of the scalar
perturbations dynamics.

\section{Noncommutative modifications}

Now we want to solve equation (40) using explicit noncommutative
forms of $\rho$, $H$,  $\omega$ and $U$ given by (20), (21) , (23)
and (39) respectively. To do this end, we need to specify the
noncommutative form of $N$ which has been appeared in equation (39).
As we have shown in Ref. [18], the noncommutative number of e-folds
is given by
$$
N=\int^{t_{f}}_{t_{i}}H dt\simeq \frac{8}{3}\pi \kappa^{2} \,{\it
\rho_{0}}\, \bigg[ \sqrt {\pi \theta}\,\,\,{\rm erf}
 \Big(\frac{1}{2}\,{\frac {t_{f}}{\sqrt {\theta}}} \Big) +\frac{1}{2}\,\sqrt {2\pi
 \theta}\,\,\,
{\rm erf} \Big( \frac{1}{2}\,{\frac {\sqrt {2}t_{f}}{ \sqrt
{\theta}}} \Big) {\lambda}^{-1} \bigg] $$
\begin{equation}
- \frac{8}{3}\,\pi \kappa^{2}\,{\it \rho_{0}}\, \bigg[ \sqrt {\pi
\theta}\,\,\, {\rm erf} \Big( \frac{1}{2}\,{\frac {t_{i}}{\sqrt
{\theta}}} \Big) +\frac{1}{2}\,\sqrt {2\pi \theta}\,\,\, {\rm erf}
\Big( \frac{1}{2}\,{\frac {\sqrt {2}t_{i}}{ \sqrt {\theta}}} \Big)
{\lambda}^{-1} \bigg]
\end{equation}
where $\rm erf(x)$ denotes the error function. By expanding the
error functions in equation (41) in series, the number of e-folds
(supposing that the universe enters the inflationary phase
immediately after the big bang,\, that is,\, $t_{i}=0$ and
$t_{f}=t$) will be given by
\begin{equation}
N\simeq \frac{8}{3}\,\pi \kappa^{2}\,{\it \rho_{0}}\, \bigg[
t-\frac{1}{12}\,{\frac {{t}^{3}}{\sqrt {\pi }{
\theta}^{\frac{3}{2}}}}+{\frac {1}{160}}\,{\frac {{t}^{5}}{\sqrt
{\pi }{\theta }^{\frac{5}{2}}}}+\frac{1}{2}\, \Big(
2\,t-\frac{1}{6}\,{\frac {\sqrt {2}{t}^{3}}{\sqrt {\pi
}{\theta}^{\frac{3}{2}}}}+\frac{1}{40}\,{\frac {\sqrt
{2}{t}^{5}}{\sqrt {\pi }{\theta} ^{\frac{5}{2}}}} \Big)
{\lambda}^{-1} \bigg].
\end{equation}
Now we can integrate equation (40) to find
$$\Phi=\frac{1}{2}(1+\omega)\frac
{\rho\,\lambda\,{\kappa}^{2}C_{0}}{2H^{2}\lambda+(1+\omega)(\kappa^{2}\rho\lambda+\kappa^{2}\rho^{2})}$$
$$-6(1+\omega)\frac{{H}^{2}\rho\,\lambda\,{{ a_{0}}}^{2} U\exp \left(
3\,\omega\,a-3\,\omega\,{a_{0}}-a+{a_{0}}
\right)}{6H^{2}\lambda+(1+\omega)(\kappa^{2}\rho\lambda+\kappa^{2}\rho^{2})}\exp{\Big(\frac{-t^2}{4\theta}\Big)}$$
\begin{equation}
+\exp\Bigg[-{\frac{1}{2}\,{\frac { 2\,{H}^{2}\lambda+(1+\omega)({
\kappa}^{2}\rho\,\lambda+{\kappa}^{2}{\rho}^{2}) }{{H}^{2}\lambda
}}}\frac{t^2}{8\theta}\Bigg].
\end{equation}
Figure (2) shows the evolution of $\Phi$ for both usual braneworld
scenario and our noncommutative setup in the high energy inflation
regime ($\rho\gg\lambda$). One should note that the subsequent
evolution of the universe after times greater than a few
$\sqrt{\theta}$ should be governed by a matter content\footnote{See
for instance [28] for particle creation in an expanding universe.}
different than the one used in equation (21) (\textit{i.e.} energy
density of the initial singularity smeared by noncommutativity). So,
the evolution of $\omega$,\, $c^{2}_{s}$ and $\Phi$ in the low
energy regime essentially should be different.
\begin{figure}[htp]
\begin{center}\includegraphics{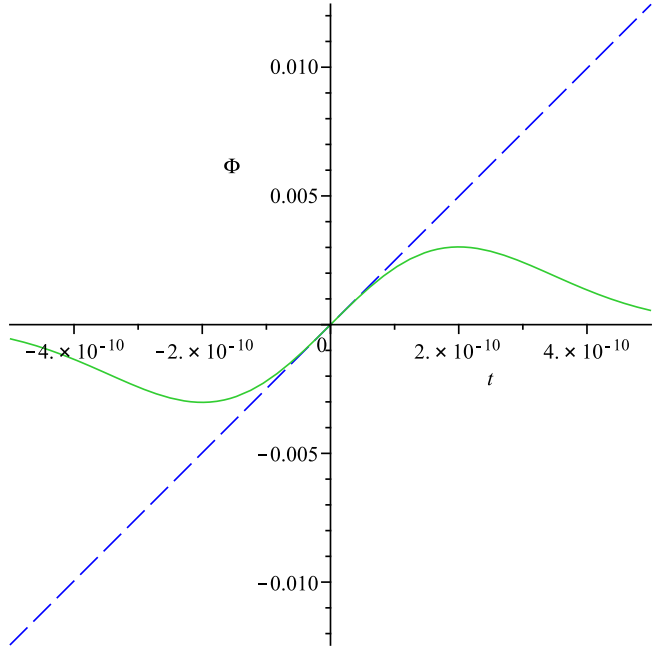} \vspace{1.5cm}
\end{center}
\vspace{6cm}
 \caption{\small {Evolution of the parameter $\Phi$ which is an analog of the Bardeen
metric potential as defined in (35) for both usual braneworld
scenario (dashed line) and our noncommutative setup (solid line)
when $\frac{\rho_{0}}{\lambda}=10^{10} $}. We assumed that no dark
radiation is present in the background geometry.}
\end{figure}
The solution (43) is valid when there is no dark radiation in the
background. If $\rho^{\varepsilon}\neq0$\,, then one should solve
the system of equations (37) and (38) using the explicit form of
$\rho^{\varepsilon}$. Generally the time dependence of
$\rho^{\varepsilon}$ for brane observer is not determined. Here we
introduce a possible candidate for this quantity: as we have
mentioned previously, the constraint from nucleosynthesis on the
value of $\rho^{\varepsilon}$ is so that
$\frac{\rho^{\varepsilon}}{\rho}\leq0.03$ at the time of
nucleosynthesis. Based on this constraint, we can assume for
instance that $\rho^{\varepsilon}$ is a small fraction of $\rho$ at
a given time. Since the time evolution of $\rho$ is determined by
(20), the time evolution of $\rho^{\varepsilon}$ can be supposed to
be
$$\rho^{\varepsilon}(t)=\frac{\delta}{32\pi^{2}\theta^{2}}e^{-t^{2}/4\theta}\, $$
where $\delta$ is a small constant and less than $0.03$. This form
of $\rho^{\varepsilon}(t)$ can be used to solve the system of
equations (37) and (38) explicitly. Nevertheless, this procedure
needs a lot of calculations with very lengthy solutions that we
ignore their
presentation here.\\
The curvature perturbation defined in metric-based perturbation
theory is
\begin{equation}
\xi={\cal R}+ {\delta\rho \over 3(\rho+p)}\,,
\end{equation}
which reduces to ${\cal R}$ on uniform density ( $\delta\rho=0$ )
hypersurfaces. If there is no dark radiation in the background
($\rho^{\varepsilon}=0$), the total curvature perturbation on large
scale is given by the following differential equation [24]
\begin{equation}
\dot{\xi}^{\rm \,eff}= \dot{\xi}^{\rm \,m}+H\left[c_{\rm
s}^2-{1\over 3}+\left({\rho+p \over \rho+ \lambda}\right)\right]
{\delta\rho^{\varepsilon} \over (\rho+p)(1+\rho/\lambda)}\,.
\end{equation}
where $\xi^{m}$ is matter perturbation which is zero for adiabattic
perturbations. Since the time variation of $\rho$,\, $H$,\, $p$,\,
$c_{s}$\, and $\delta\rho^{\varepsilon}$ are given by equations
(20), (21), (22), (24) and (39) respectively, we can obtain the time
evolution of the curvature perturbation explicitly as follows
$$\xi^{\rm \,eff}={\frac {1}{96}}\frac{\,{\rm \,Ei} \left( 1,{\frac
{3{t}^{2}}{8\theta}} \right)} {{\pi
}^{2}{\theta}^{2}{\lambda}}-\frac{1}{48}\,\frac{{{\rm e}^{- {\frac
{{3t}^{2}}{8\theta}}}}}{{\pi }^{2}{\theta}^{2}{\lambda}}+\frac{1}{3}
\,{\rm \,Ei} \left( 1,{\frac {{t}^{2}}{4\theta}} \right)
-\frac{2}{3}\,{ {\rm e}^{-{\frac {{t}^{2}}{4\theta}}}}$$
\begin{equation}
-{\frac {1}{768}}\, {{\rm \rm erf}\left({\frac {t}{2\sqrt
{\theta}}}\right)}{\lambda}^{-1} {\pi }^{-3}{\theta}^{-3}{\frac
{1}{\sqrt {\theta\,\pi }}}-\frac{1}{24}\, {{\rm
erf}\left(\frac{1}{4}\,{\frac {\sqrt {2}t}{\sqrt {\theta}}}\right)}
\sqrt {2}{\theta}^{-1}{\pi }^{-1}{\frac {1}{\sqrt {\theta\,\pi }}}
\end{equation}
Where   ${\rm \,Ei}(a,z)$ is the exponential integral defined as
${\rm \,Ei}(a,z)=z^{a-1}\Gamma(1-a,z)$. Figure $3$ shows the
evolution of ${\xi}^{\rm \,eff}$ versus the cosmic time for both
commutative and noncommutative brane inflation. For the commutative
braneworld inflation, we have considered the chaotic inflation with
$V(\phi)=\frac{1}{2}m^{2}\phi^{2}$ and we obtained the form of
$\phi$ using its relation with $\rho$. The amplitude of
perturbations in the commutative regime decays faster than the
noncommutative regime with the same parameter values. As a result we
need smaller number of e-folds in the noncommutative regime to have
a successful braneworld inflation.

\begin{figure}[htp]
\begin{center}\includegraphics{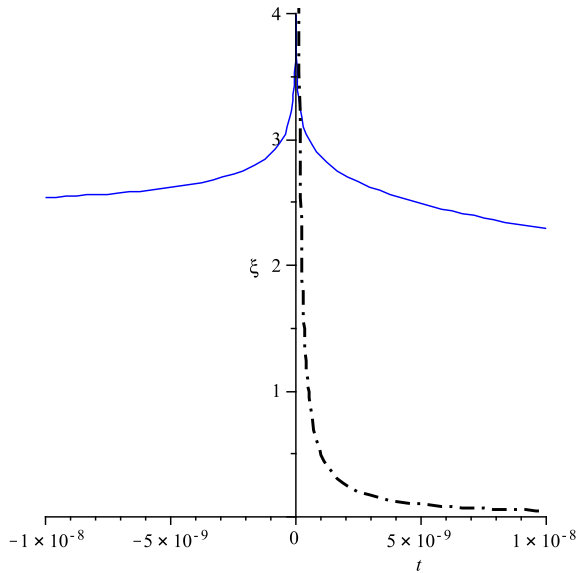} \vspace{1.5cm}
\end{center}
\vspace{6cm}
 \caption{\small { a) Evolution of the parameter $\xi$ in
 commutative brane inflation ( dashed line).
 b) Evolution of the parameter $\xi$
 defined in (44)( with the same parameters values as in figure (2))
 when $\frac{\rho_{0}}{\lambda}=10^{10} $} and no dark
 radiation is present in the background geometry (solid lines). }
\end{figure}

\section{Conclusion}
Spacetime noncommutativity as a trans-Planckian effect, essentially
could have some observable effects on the cosmic microwave
background radiation. In this respect, it is desirable to study an
inflation scenario within a noncommutative background. Recently we
have shown the possibility of realization of a non-singular,
bouncing, early time cosmology in a noncommutative braneworld
scenario [18]. In that work, using the smeared, coherent state
picture of the spacetime noncommutativity, we have constructed a
braneworld inflation that has the potential to support the scale
invariant scalar perturbations. Here, following our previous work,
we have studied the time evolution of the perturbations in this
noncommutative braneworld setup. We have neglected the contribution
of the dark radiation term ( originating in the bulk Weyl tensor) in
the background geometry to have a closed set of equations on the
brane. However, the contribution of this term in the evolution of
perturbations on the brane are taken into account. In this way, by
studying the effective quantities ( such as the effective equation
of state and speed of sound ) we have realized the possibility of a
phantom evolution in the early, inflationary stage of the universe
history. Our analysis of the perturbations on the brane shows that
the amplitude of perturbations in the commutative regime decays
faster than the noncommutative regime with the same parameter
values. As a result we need smaller number of e-folds in the
noncommutative regime to have a successful braneworld inflation.\\

{\bf Acknowledgment}\\
This work has been supported partially by Research Institute for
Astronomy and Astrophysics of Maragha, IRAN.

\end{document}